\documentclass[aps,pre,floatfix]{revtex4}
\usepackage{graphicx}% Include figure files
\usepackage{dcolumn}% Align table columns on decimal point
\usepackage{bm}% bold math
\usepackage{amssymb}
\usepackage{amsmath}
\usepackage{epsfig}
\def\LM#1#2{\left|\begin{array}{l}{#1}\\[1ex]{#2}\end{array}\right.}
\def\toi{\to \infty}
\begin{document}
\title{Trapping reactions with subdiffusive traps and particles}
\author{S. B. Yuste$^{1}$ and Katja Lindenberg$^{2}$}
%\skiplinehalf
%\supit{a}
\affiliation{$^{(1)}$
Departamento de F\'{\i}sica, Universidad de Extremadura,
E-06071 Badajoz, Spain\\
$^{(2)}$Department of Chemistry and Biochemistry, and Institute for
Nonlinear Science,
University of California San Diego, 9500 Gilman Drive, La Jolla, CA
92093-0340, USA}
%\authorinfo{Further author information: Send correspondence to
%S.B.Y.\\S.B.Y.: E-mail: santos@unex.es, Telephone: +34 924 289 529\\
%K.L. E-mail: klindenberg@ucsd.edu, Telephone: +1 858 534 3285}
%\maketitle
\begin{abstract}
Reaction dynamics involving subdiffusive species is an interesting topic with
only few known results, especially when the motion of different species
is characterized by different anomalous diffusion exponents.  Here we
study the reaction dynamics of a (sub)diffusive particle surrounded by
a sea of (sub)diffusive traps in one dimension. Under some reasonable
assumptions we find rigorous results for the asymptotic survival
probability of the particle in most cases, but have not succeeded in
doing so for a particle that diffuses normally while the anomalous
diffusion exponent of the traps is smaller than 2/3.
\end{abstract}
%\keywords{Subdiffusion, diffusion, reactions, mobile traps, survival probability}
\maketitle
\section{INTRODUCTION}
\label{sec:intro}
In the traditional version of the trapping problem, a normal diffusive
(Brownian) particle ($A$) wanders in a medium doped at random with {\em
static} traps ($B$), and disappears when they meet, $A+B\rightarrow B$.
The quantity of interest is the survival probability $P(t)$ of the
particle $A$.  This problem dates back to Smoluchowski's theory of
reaction rates at the beginning of last century,
and is one of the most widely investigated and applied problems of
non-equilibrium statistical
mechanics~\cite{Hughes1,Hughes2,Weiss,HollanderWeiss,ShDba,AvrahamHavlinDifuReacBook}.
An important variation of the basic trapping problem, in which a
diffusive particle wanders in a medium in which the traps are also
\emph{diffusive}, has been the subject of intense research
since the seminal work of Toussaint and Wilczek~\cite{ToussaintWilczekPRL}.

The principal quantity of interest in the trapping problem is the
survival probability $P(t)$ of the $A$ particles.  From this survival
probability one is able to calculate essentially all other quantities of
practical interest.  Yet this probability is usually difficult to calculate,
and the few instances in which it has been obtained are considered
landmark contributions.  In 1988, Bramson and
Lebowitz~\cite{BramsonLebowitz1,BramsonLebowitz2} proved
rigorously that the long-time survival probability $P(t)$ of a
particle diffusing in a one-dimensional medium doped with
diffusive traps decays as $P(t)\sim \exp(-\lambda t^{1/2})$,
$\lambda$ being an undetermined parameter. The evaluation of this
constant proved elusive for many years, engendering much confusion and
proposed solutions that were mutually contradictory. Finally,
quite recently Bray and
Blythe~\cite{BrayBlythePRL,BlytheBrayPRE} proved in a
simple and elegant way, assuming the so-called ``Pascal principle,"
that the survival probability $P(t)$ of a
diffusing particle with diffusion coefficient $D'$ in a
$d$-dimensional medium with $d\le 2$ in which the traps are
also diffusive with diffusion coefficient $D$ is
\emph{independent} of $D'$ for long times, and coincides with the
survival probability of an immobile target ($D'=0$) in the
presence of a density $\rho$ of diffusive traps. In particular, in a
one-dimensional medium $P(t)\sim \exp(-4\rho (D
t/\pi)^{1/2})$.  Bray and Blythe obtained their results by calculating
an upper and a lower bound for the survival probability that converge to
one another asymptotically.
Some (but not all) of the bounding
results of Bray and Blythe have been extended by Oshanin et
al.~\cite{OshaninEtAlPRE} to systems where the traps
perform a compact exploration of the space, i. e.,
where the fractal dimension $d_w$ of the trajectories of the
traps is greater than the dimension $d$ of the space.

The purpose of this work is to extend the procedure and results of
Bray and Blythe~\cite{BrayBlythePRL,BlytheBrayPRE}, which are
valid for a Brownian diffusive particle and Brownian diffusive
traps, to situations in which the particle and traps exhibit
anomalous diffusion, in particular, subdiffusion. The usual
characterization of anomalous
diffusion of a particle is through its mean squared displacement
$x(t)$ for large $t$:
\begin{equation}
\left< x^2(t)\right> \sim \frac{2K_\gamma}{\Gamma(1+\gamma)}
t^\gamma . \label{meansquaredispl}
\end{equation}
Here $K_\gamma$ is the (generalized) diffusion constant and
$\gamma$ is the (anomalous) diffusion exponent. Ordinary Brownian
diffusion ($\gamma=1$, $K_1\equiv D$) follows Fick's
second law, $\left< x^2(t)\right> \propto t$. The process is
called sudiffusive when $0<\gamma<1$. Subiffusive process are
ubiquitous in
nature~\cite{MetKlaPhysRep,BouchaudPhysRep90,Kosztolowicz,SubdifuRandPot1,SubdifuRandPot2,SubdifuRandPot3,KantorCM,ChenDeemPRE,GalloRovere03},
and are particularly useful for understanding transport in complex
systems~\cite{ShDba,BouchaudJPI,ArousPRL02,BertinPRE03}.

Two main approaches have been used to study subdiffusive
processes. The older is based on the continuous time random
walk (CTRW) theory with waiting-time distributions between steps that
have broad long-time tails and consequently infinite moments,
$\psi(t)\sim t^{-1-\gamma}$ for $t\to \infty$ with $0<\gamma<1$.
Another approach is based on the fractional diffusion equation,
which describes the evolution of the probability density $P(x,t)$
of finding the particle at position $x$ at time $t$ by means of
the fractional partial differential equation (in one
dimension)~\cite{MetKlaPhysRep,SchWysJMP}
\begin{equation}
\frac{\partial }{\partial t} P(x,t)= K_\gamma
~_{0}\,D_{t}^{1-\gamma } \frac{\partial^2}{\partial x^2} P(x,t)
\label{Pfracdifu}
\end{equation}
where $K_\gamma$ is the generalized diffusion coefficient that
appears in Eq.~(\ref{meansquaredispl}) and
$~_{0}\,D_{t}^{1-\gamma } $ is the Riemann-Liouville
operator,
\begin{equation}
~_{0}\,D_{t}^{1-\gamma } P(x,t)=\frac{1}{\Gamma(\gamma)}
\frac{\partial}{\partial t} \int_0^t d\tau
\frac{P(x,\tau)}{(t-\tau)^{1-\gamma}}.
\end{equation}
Here we implement the latter approach to
study the one-dimensional trapping problem in the
long-time regime for subdiffusive (or diffusive) particles that
move among a distribution of \emph{non-static}
traps. The traps can be either subdiffusive or (Brownian)
diffusive.  For this purpose, we generalize the
ideas of Bray and Blythe~\cite{BrayBlythePRL,BlytheBrayPRE}.

The problem considered in this paper is a special case of
a broad class of reaction-\emph{subdiffusion} processes that have been
studied over the past decades using one or the other of the established
formalisms.  Using the CTRW formalism, Blumen
et al.~\cite{Klafter,Zumofen,BluKlaZuOptical} considered a variety of
reactions including the trapping problem
$A+T(\text{static})\to T(\text{static})$,
the target problem $A\text{(static)}+T\to T$,
and the bimolecular reactions $A+A\to \emptyset$ and $A+B\to \emptyset$.
The moving particles were modeled as continuous-time random walkers with
long-tailed waiting-time densities.  Quite recently, Sung and
Silbey~\cite{SungSilbeyPRL03} have used the CTRW model to study the
dynamics of particles that react at a boundary.  A CTRW approach has
also been applied by
Seki et al.~\cite{SekiJCP1s03,SekiJCP1s03} to study the kinetics of
the recombination
reaction in subdiffusive media. However, Seki et al. went further and,
from the
CTRW model, derived a fractional reaction-diffusion equation for
the geminate recombination problem. Sung et al.~\cite{SungJCP02}
directly addressed
this problem from a fractional diffusion equation approach, but
some of their assumptions and results
disagree with those of Seki et al.
The fractional diffusion approach
has recently been used to get exact solutions for two types of
one-dimensional
trapping problems: the so called one-sided problem, in which all the
traps lie
on one side of the particle, and the two-sided problem, in which the
traps are located on
both sides of the particle (this is the traditional or standard
version of
the trapping problem)~\cite{YusteAcedoSubTrap}.  These recent
articles~\cite{SungSilbeyPRL03,YusteAcedoSubTrap} share the
simplifying characteristic that the reaction takes place between a
static particle (or fixed boundary) and a subdiffusive particle.
The present paper differs from these
in that all the reacting particles (including traps) are
(sub)diffusive and,
moreover, the diffusion constant and the anomalous diffusion exponent of
each species may be different.  The fractional-diffusion approach
has already been employed to study bimolecular reactions between
subdiffusive particles.  In
particular, the annihilation $A+A \to \emptyset$ and coagulation $A+A
\to A$ of subdiffusive particles was
studied~\cite{YusteKatja1,YusteKatja2} by
means of a fractional generalization of the interparticle distribution
function method~\cite{AvrahamHavlinDifuReacBook}. The evolution of
reaction-subdiffusion fronts
for $A+B\to C$ reactions, where both $A$ and $B$ move
subdiffusively, is also
amenable to analysis by means of the fractional diffusion
approach~\cite{YusAceLinSubFront}. Other recent work on fractional
diffusion and CTRW models
of subdiffusive reacting particles can be found in a number of
references~\cite{SubdifuReactOtros1,SubdifuReactOtros2,SubdifuReactOtros3,SubdifuReactOtros4}.

In some cases, asymptotic anomalous diffusion behavior can be found from
corresponding results for normal diffusion with the simple replacement
of $t$ by $t^\gamma$ wherever $t$ appears.  This can be understood from
a CTRW perpective because
the average number of jumps $n$ made by a subdiffusive walker up to
time $t$
scales as $\langle n \rangle \sim t^\gamma$ and, in many instances the
number of jumps is the relevant factor that explains the
behavior of the system.
The simple replacement result is evidence
of ``subordination''~\cite{BluKlaZuOptical}.  Examples of
this phenomenon are given in Secs. 5 and 7.2 of the preceding reference.
However, there are other instances where
the behavior of subdiffusive systems cannot be found in this way.
A simple example is the survival probability of subdiffusive
particles in
the trapping problem, see Sec. 5 of the preceding reference. In
particular, for
systems where each species has a different anomalous diffusion
exponent, such a replacement becomes ambiguous.
This is the case for the problem considered here.

Bray and Blythe obtained the asymptotic survival probability of a diffusing
particle in a sea of diffusing traps by calculating an
upper and lower bound that converge asymptotically. We follow their
procedure for subdiffusive particle and traps, with partial success.
While it is possible to obtain convergent bounds in most anomalous
diffusion exponent regimes, this procedure does not work in all regimes.
In particular, the bounding procedure encounters difficulties when the
particle $A$ diffuses normally and the traps are ``too slow."
In Sec.~\ref{Sec:upperbound} we calculate
the upper bound of the survival probability,
and in Sec.~\ref{Sec:lowerbound} the lower bound.
The survival probability is established, when possible, in
Sec.~\ref{Sec:survival}.  Section~\ref{Sec:panorama} and the Appendix
presents a compendiary of results and some comments on open problems.

\section{UPPER BOUND FOR THE SURVIVAL PROBABILITY}
\label{Sec:upperbound}
The ``Pascal principle'' of random walks says that the best
survival strategy for a random walker $A$ surrounded by a sea of
trapping random walkers $B$ is to stand still.  This assumption was
adopted in one dimension by Bray and
Blythe~\cite{BrayBlythePRL,BlytheBrayPRE}, and proved by
Bray, Majumdar and Blythe~\cite{BrayMajumBlythePRE} for
$d\leq 2$. Almost simultaneously,
Moreau et al.~\cite{MoreauEtAlCondMat} proved
the Pascal principle for a rather general class of random walks on
$d$-dimensional lattices. Although the case in which both
particle and traps perform subdiffusive random walks was not considered,
the Pascal principle
is an intuitively plausible (and most likely provable) result for
this case as well, and we will simply assume that it is applicable.

The Pascal principle thus says that the survival probability $P_U(t)$ of a
static particle $A$ surrounded by a random (Poisson) distribution of
randomly walking traps (the ``target problem") is an upper bound for
the survival
probability $P(t)$ of a (sub)diffusing particle $A$.  We proceed to
calculate this upper bound for the subdiffusive target problem.
This problem has been considered~\cite{BluKlaZuOptical} by
means of the CTRW model and, for the three-dimensional
case~\cite{SungJCP02} by means of a fractional diffusion approach. Here we
calculate $P_U(t)$ by generalizing the approach of
Bray and Blythe~\cite{BrayBlythePRL,BlytheBrayPRE} to the subdiffusive
case.

Consider a target of size $2L$ centered at the origen,
and let $Q_1(t|y)$ be the probability that the trap initially placed at
$y>L$ has not reached the
target $A$ at $y=L$ by time $t$. Then~\cite{MetKlaBoundary}, in terms of
the Fox's $H$ function,
\begin{equation}\label{}
Q_1(t|y)=1-H^{10}_{11}\left[\frac{y-L}{\sqrt{K_\gamma t^\gamma}}
    \LM{(1 ,\gamma/2)}{(0,1)}   \right]\equiv
1-H\left[\frac{y-L}{\sqrt{K_\gamma t^\gamma}} \right].
\end{equation}
For $\gamma\rightarrow 1$ the Fox's $H$ function becomes the
complementary error function (we set $K_1 \equiv D$),
and the ordinary Brownian motion result is recovered,
\begin{equation}\label{Brownian}
Q_1(t|y) = 1- \text{erfc}\left(\frac{y-L}{\sqrt{4 D t}}\right), \qquad
\gamma=1.
\end{equation}
Next consider $N$
independently diffusing traps that, at $t=0$, are placed  at
random in the interval $L\leq y \leq L+R$. Here and henceforth $2R$ is the
size of the system, which we will take to infinity at appropriate points
in the calculations. The probability
$Q_N(t)$ that the stationary target $A$ has survived up to time $t$ is
\begin{align}\label{}
Q_N(t)&=\prod_{i=1}^N \frac{1}{R}\int_L^{L+R} dy_i
\left\{1-H\left[\frac{y_i-L}{\sqrt{K_\gamma
t^\gamma}}\right]\right\}
=\left\{1-\frac{1}{R}\int_L^{L+R} dy
H\left[\frac{y-L}{\sqrt{K_\gamma t^\gamma}} \right]\right\}^N,
\end{align}
or, in terms of the density $\rho=N/R$ of traps,
\begin{align}\label{}
Q_\infty(t)&=\lim_{R\rightarrow \infty}
\left\{1-\frac{1}{R}\int_L^{L+R} dy
H\left[\frac{y}{\sqrt{K_\gamma t^\gamma}} \right]\right\}^{\rho
R}= \exp\left\{-\rho \sqrt{K_\gamma t^\gamma} \int_0^{\infty}
dz\, H\left[z \right]\right\}.
\end{align}
We need to evaluate the integral
\begin{align}\label{}
I_\gamma&= \int_0^{\infty} dz\,H^{10}_{11}\left[z
    \LM{(1 ,\gamma/2)}{(0,1)}   \right],
\end{align}
which can be done from the properties of the Fox's $H$
function~\cite{MathaiSaxena}. One finds that
\begin{equation}\label{}
H^{10}_{11}\left[z
    \LM{(1 ,\gamma/2)}{(0,1)}\right]=\frac{d}{dz} \,H^{10}_{11}\left[z
    \LM{(1+\gamma/2 ,\gamma/2)}{(0,1)}\right].
\end{equation}
But
\begin{equation}\label{}
H^{10}_{11}\left[\infty\LM{(1+\gamma/2,\gamma/2)}{(0,1)}\right]=0
\end{equation}
and
\begin{equation}\label{}
H^{10}_{11}\left[0\LM{(1+\gamma/2,\gamma/2)}{(0,1)}
\right]=\frac{1}{\Gamma(1+\gamma/2)},
\end{equation}
so that
\begin{equation}\label{}
I_\gamma=\frac{1}{\Gamma(1+\gamma/2)} .
\end{equation}
Therefore,
\begin{equation}\label{eq8}
Q_\infty(t)=\exp\left[-\frac{\sqrt{\rho^2 K_\gamma
t^\gamma}}{\Gamma(1+\gamma/2)}  \right].
\end{equation}
This is the survival probability of the target when all the
traps are located to its right.  When the traps are located on
both sides of the target, the survival probability of the target
is the square of Eq.~\eqref{eq8}:
\begin{equation}\label{eq8b}
P_U(t)=Q_\infty^2(t)=\exp\left[-\frac{2\sqrt{\rho^2 K_\gamma
t^\gamma}}{\Gamma(1+\gamma/2)}  \right] .
\end{equation}
This is the \emph{upper bound} on the survival probability of the moving
particle.

Incidentally, as is well known, the survival probability for the
target problem is related to the distinct number of sites $S(t)$
visited by a trap up to time $t$:~\cite{BluKlaZuOptical,BenichouPRE00}
\begin{equation}\label{eq8c}
P_U(t)=e^{-\rho \langle S(t)\rangle }.
\end{equation}
Comparing this expression with Eq.~\eqref{eq8b}, one
finds that the average value $\langle S(t)\rangle$ of the territory explored
up to time $t$ by a subdiffusive walker with
generalized diffusion coefficient
$K_\gamma$ and anomalous diffusion exponent $\gamma$ is
\begin{equation}\label{}
\langle S(t)\rangle \sim \frac{2\sqrt{ K_\gamma
t^\gamma}}{\Gamma(1+\gamma/2)}.
\end{equation}
This result agrees with that found by Yuste and Acedo~\cite{YusteAcedoSubTrap}
using a different approach.

\section{LOWER BOUND FOR THE SURVIVAL PROBABILITY}
\label{Sec:lowerbound}
Let $P_L(t)$ be the  probability that the particle $A$ (which is
now allowed to move) remains inside a box of
size $L$ and that all the traps remain outside this box until
time $t$. When this
happens, the particle $A$ survives. It is clear that $P_L(t)$ is a
lower bound for the survival probability $P(t)$ of interest here
because there exist many other ways in which the
particle $A$ may survive. The probability $P_L(t)$ is itself the product of
three probabilities:
\begin{enumerate}
\item The probability $Q_1$ that at $t=0$ the box of size $L$ contains no
traps:
\begin{equation}\label{notraps}
Q_1=e^{-\rho L}.
\end{equation}
\item The probability $Q_2$ that no traps enter the box of size $L$ up to time
$t$:
\begin{equation}\label{}
Q_2=\exp\left[-\frac{2}{\Gamma(1+\gamma/2)} \sqrt{\rho^2 K_\gamma
t^\gamma}\right] =P_U(t).
\end{equation}
\item The probability $Q_3$ that the particle has not left the box of
size $L$ up to time $t$.  We proceed to evaluate this quantity.
\end{enumerate}

Let $W(x,t)$ be the probability of finding the particle $A$ at
position  $x$ at time $t$ if it was at position $x=0$ at time
$t=0$ and there are absorbing boundaries at $x=-L/2$ and $x=L/2$.
Solving the fractional diffusion equation by means of separation
of variables~\cite{MetKlaPhysRep} one finds
\begin{align}\label{}
W(x,t) &=\frac{2}{L}\sum_{n=0}^\infty (-1)^n \sin \frac{(2n+1)\pi
(x+L/2)}{L} E_{\gamma'}\left(-K'_{\gamma'} (2n+1)^2\pi^2
t^{\gamma'}/L^2\right),
\end{align}
where $K'$ and $\gamma'$ are the generalized diffusion constant and the
anomalous diffusion exponent of the particle $A$.  Therefore,
\begin{align}
Q_3= \int_{-L/2}^{L/2} W(x,t) dx &=\frac{4}{\pi}\sum_{n=0}^\infty
\frac{(-1)^n}{2n+1}E_{\gamma'}\left[-K'_{\gamma'} (2n+1)^2\pi^2
t^{\gamma'}/L^2\right]. \label{Q3a}
\end{align}

Next we distinguish two cases in the handling of the sum in
Eq.~\eqref{Q3a}: first we deal with a
subdiffusive particle, and subsequently with an ordinary diffusive
particle.
In the subdiffusive case, we note that
for large arguments ($z\gg 1$) the Mittag-Leffler function has the
expansion
\begin{equation}\label{}
E_{\gamma'}(-z)=\sum_{m=1}^\infty \frac{(-1)^{m+1}
}{\Gamma(1-\gamma' m) } \,z^{-m}
\end{equation}
so that
\begin{align}\label{}
Q_3= & \frac{4}{\pi} \sum_{m=1}^\infty
\frac{(-1)^{m+1}L^{2m}}{\Gamma(1-\gamma' m ) \left[\pi^2
K'_{\gamma'} t^{\gamma'}\right]^m} \sum_{n=0}^\infty
\frac{(-1)^n}{(2n+1)^{m+1}}.
\end{align}
Therefore, for $t \rightarrow \infty$ one finds
\begin{align}\label{}
Q_3=  & \frac{1}{8\Gamma(1-\gamma')} \frac{L^{2}}{K'_{\gamma'}
t^{\gamma'} } +O\left(\frac{L^{2}}{K'_{\gamma'} t^{\gamma'}
}\right)^2.
\end{align}
Consequently, a lower bound on the survival probability of the particle
$A$ is
\begin{equation}\label{}
P_L(t)=Q_1Q_2Q_3= e^{-\rho L}
\exp\left[-\frac{2}{\Gamma(1+\gamma/2)} \sqrt{\rho^2 K_\gamma
t^\gamma}\right]  \frac{1}{8\Gamma(1-\gamma')}
\frac{L^{2}}{K'_{\gamma'} t^{\gamma'} }
\left[1+O\left(\frac{L^{2}}{K'_{\gamma'} t^{\gamma'}
}\right)\right].
\end{equation}
It can easily be ascertained that this expression is maximal
when $L=L^*\equiv 2/\rho$, i.e., $P_L(t)\leq P_{L^*}(t)$ with
\begin{equation}\label{eq:bound1}
P_{L^*}(t)= \frac{e^{-2}}{8\Gamma(1-\gamma')}
\left(\frac{2}{\rho}\right)^2 \frac{1}{K'_{\gamma'} t^{\gamma'} }
\exp\left[-\frac{2}{\Gamma(1+\gamma/2)} \sqrt{\rho^2 K_\gamma
t^\gamma}\right] \left[1+O\left(\frac{1}{\rho^2K'_{\gamma'}
t^{\gamma'} }\right)\right].
\end{equation}
This is then our best lower bound for the survival probability $P(t)$ of
a subdiffusive particle.

When the particle $A$ diffuses normally, Eq.~\eqref{Q3a} becomes
\begin{align}\label{}
Q_3= &\frac{4}{\pi}\sum_{n=0}^\infty
\frac{(-1)^n}{2n+1}\exp\left[-D'(2n+1)^2\pi^2 t/L^2\right],
\end{align}
with $D'\equiv K'_1$.  For long times~\cite{BrayBlythePRL,BlytheBrayPRE}
\begin{align}\label{}
Q_3\sim &\frac{4}{\pi}\exp\left[-D'\pi^2 t/L^2\right], \qquad t\gg 1
\end{align}
so that
\begin{equation}\label{}
P_L(t)=Q_1Q_2Q_3\sim \frac{4}{\pi} e^{-\rho L}
\exp\left[-\frac{2}{\Gamma(1+\gamma/2)} \sqrt{\rho^2 K_\gamma t^\gamma}\right]
\exp\left[-D'\pi^2 t/L^2\right]
\end{equation}
for $t\gg 1$.  This lower bound can be again be maximized by
optimizing the value of $L$.  The optimal value
is~\cite{BrayBlythePRL,BlytheBrayPRE}
$L^*=\left(2\pi^2 D' t/\rho\right)^{1/3}$,
so that
\begin{equation}\label{eq:dominant}
P_L(t)\leq P_{L^*}(t)= \frac{4}{\pi} \exp\left[-\frac{2\sqrt{\rho^2 K_\gamma
t^\gamma}}{\Gamma(1+\gamma/2)} -3(\pi^2 \rho^2 D't/4)^{1/3}\right].
\end{equation}
Note that
the dominant term inside the bracket depends on the value of $\gamma$,
the anomalous diffusion exponent for the traps. We
distinguish three cases:
\begin{enumerate}
\item  Traps with $2/3<\gamma\leq 1$. In this case, for $t\gg 1$,
\begin{equation}\label{}
\frac{2\sqrt{\rho^2 K_\gamma t^\gamma}}{\Gamma(1+\gamma/2)} \gg 3(\pi^2 \rho^2
D't/4)^{1/3}
\end{equation}
so that
\begin{equation}
P_{L^*}(t)= \frac{4}{\pi} \exp\left[-\frac{2\sqrt{\rho^2 K_\gamma
t^\gamma}}{\Gamma(1+\gamma/2)} \right].
\end{equation}
\item Traps with $\gamma=2/3$. Now
\begin{equation}\label{}
P_L(t)\leq P_{L^*}(t)= \frac{4}{\pi} \exp\left[-\left(\frac{2\sqrt{\rho^2
K_\gamma }}{\Gamma(4/3)} -3(\pi^2 \rho^2 D'/4)^{1/3}\right)
t^{1/3}\right],
\end{equation}
that is, the second contribution in the exponent in
Eq.~\eqref{eq:dominant} is of the same order as
the first and must thus be retained.
\item
\label{caso3} Traps with $0<\gamma<2/3$. Now the second term in the
exponent of Eq.~\eqref{eq:dominant} is dominant:
\begin{equation}
P_{L^*}(t)= \frac{4}{\pi} \exp\left[ -3(\pi^2 \rho^2 D't/4)^{1/3}\right].
\end{equation}
\end{enumerate}

In the next section we examine our upper and lower bound results to
establish the behavior of the survival probability of $A$ whenever
possible.

\section{SURVIVAL PROBABILITY}
\label{Sec:survival}
We now combine our upper and lower bound results.
Recall that the label and exponent $\gamma$
is associated with the traps and $\gamma'$ is associated with the
particle $A$.  The upper bound on the survival probability is in all
cases given in Eq.~\eqref{eq8b}, but the lower bound depends on the anomalous
diffusion exponent of the particle. We distinguish the following cases:
\begin{enumerate}
\item Subdiffusive particle ($0<\gamma'< 1$) and diffusive or
subdiffusive traps ($0<\gamma \leq 1$).  The lower bound is given in
Eq.~\eqref{eq:bound1}, so that $P_{L^*}(t)\leq P(t)\leq P_U(t)$ leads to
\begin{align}
\frac{2}{\Gamma(1+\gamma/2)} \leq  -\frac{\ln P(t)}{\sqrt{\rho^2 K_\gamma
t^\gamma}} &\leq  \frac{2}{\Gamma(1+\gamma/2)} + \frac{2\ln\left[\sqrt{\rho^2
K'_{\gamma'} t^{\gamma'}}\right]+2+\ln\left[2\Gamma(1-\gamma')\right]
}{\sqrt{\rho^2 K_\gamma t^\gamma}}  \nonumber\\
& ~~~~+O\left(\frac{(\rho^2 K_\gamma t^\gamma)^{-1/2}}
{\rho^2K'_{\gamma'} t^{\gamma'}
}\right).
\end{align}
For $t\rightarrow \infty$,
$\ln\left[\sqrt{\rho^2 K'_{\gamma'} t^{\gamma'}}\right]\ll
\sqrt{\rho^2 K_\gamma t^\gamma}$
and the upper and lower bounds converge asymptotically.
We therefore arrive at the explicit asymptotic survival probability
\begin{equation}\label{Ptgral}
P(t)\sim \exp\left[-\frac{2}{\Gamma(1+\gamma/2)} \sqrt{\rho^2
K_\gamma t^\gamma} \right]
\end{equation}
for $0<\gamma\leq 1$ and $0<\gamma'< 1$. Note that for $\gamma=1$ we
recover the normal diffusive result obtained earlier~\cite{BrayBlythePRL}.
A noteworthy result here is that the survival
probability depends only on the exponent $\gamma$ that characterizes the
traps and not on $\gamma'$ that characterizes the particle.  This is
interesting vis a vis the subordination issue.

\item Diffusive particle ($\gamma'=1$) and subdiffusive traps with
$2/3<\gamma\leq 1$. The bounds here are
\begin{align}
\frac{2}{\Gamma(1+\gamma/2)} &\leq -\frac{\ln P(t)}{\sqrt{\rho^2 K_\gamma
t^\gamma}} \leq \frac{2}{\Gamma(1+\gamma/2)} + 3
\left(\frac{\pi}{2\rho}\right)^{2/3} \frac{D'^{1/3}}{
 K_\gamma^{1/2} } t^{1/3-\gamma/2},
\label{boundsga1}
\end{align}
and the asymptotic survival probability is again given by Eq.~\eqref{Ptgral}.

\item Diffusive particle ($\gamma'=1$) and subdiffusive traps with
$\gamma=2/3$ (marginal case).
Now $P_{L^*}(t)\leq P(t)\leq P_U(t)$ leads to the more ambiguous
inequalities
\begin{align}
\frac{2}{\Gamma(4/3)} &\leq -\frac{\ln P(t)}{\sqrt{\rho^2 K_\gamma
t^\gamma}}
\leq \frac{2}{\Gamma(4/3)} + 3 \left(\frac{\pi}{2\rho}\right)^{2/3}
\frac{D'^{1/3}}{
 K_\gamma^{1/2}}.
\end{align}
The bounding procedure is therefore not able to predict the value of the
prefactor $\lambda$ in $P(t)=\exp(-\lambda t^{1/3})$, but the
asymptotic behavior $-\ln P(t) \propto t^{1/3}$ is evident.

\item Diffusive particle ($\gamma'=1$) and subdiffusive traps with
$0<\gamma<2/3$.
The bounds here are also given by Eq.~\eqref{boundsga1}, so that the bounding procedure is not able to determine the asymptotic behavior of $P(t)$ at all for this case. We are not even able to assert
the asymptotic stretched exponential
form $P(t)\sim \exp\left(-\lambda t^\beta \right)$.
\end{enumerate}

\section{PANORAMA AND DISCUSSION}
\label{Sec:panorama}
Bray and Blythe~\cite{BrayBlythePRL,BlytheBrayPRE} have calculated
the asymptotic survival probability of
a diffusive particle $A$ in a randomly distributed sea of
diffusive traps $B$ in one
dimension, and have determined the precise value of the coefficient
$\lambda$ in the classic result $P(t)\sim \exp \left( -\lambda
t^{1/2}\right)$ first obtained by Bramson and
Lebowitz~\cite{BramsonLebowitz1,BramsonLebowitz2}.  We have
attempted to generalize this result to the case where one or both of the
species move subdiffusively.  Our particle $A$ is
characterized by the anomalous diffusion exponent $\gamma'$ and
generalized diffusion coefficient $K_{\gamma'}$, and the
traps by $\gamma$ and $K_\gamma$. These may be the first results in the
literature involving two subdiffusive species with different anomalous
diffusion exponents.

When both species are subdiffusive ($\gamma$ and $\gamma'$
both smaller than unity), the
survival probability is independent of $\gamma'$ and determined
entirely by the subdiffusive properties of the traps, cf.
Eq.~\eqref{Ptgral}.  When the particle
moves diffusively ($\gamma'=1$), on the other hand, we are unable to
unequivocally determine the coefficient $\lambda$ for all cases using
this procedure.  If the traps move sufficiently rapidly
($2/3<\gamma\leq 1$) then the result Eq.~\eqref{Ptgral} is still valid.
Note that this reduces to the Bray and Blythe result when $\gamma=1$.
The case $\gamma=2/3$ is marginal in the sense that we can establish the
behavior $P(t)\sim \exp \left( -\lambda t^{1/3}\right)$, but are not
able to determine the constant $\lambda$.
Note that this particular time dependence of the survival probability is
the same as the classic result for the survival probability of a
diffusive particle in a sea of immobile
traps~\cite{donsker,movaghar,zumofen2,kang}.
If the traps are too
slow (``extremely subdiffusive"), $0<\gamma<2/3$, it is no longer possible
to determine even the time dependence of the survival probability on the
basis of this approach.

We thus find that in so far as one can even think of some sort of
subordination principle (and whether such thinking is appropriate
is debatable), it is determined by the behavior of the
traps, i.e., by the replacement of $t$ by $t^\gamma$.  Even in the range
of exponents where this is possible, it is only possible for the main
asympotic contribution to $P(t)$ but not for the correction
terms to the leading asymptotic term.

It is interesting to note that the value $L^*=2/\rho$ that maximizes the
lower bound of the survival probability for a subdiffusive particle
$A$ does not grow with time.  This implies that
finite particle size effects could become relevant with increasing
density $\rho$.  This is completely different from the case of a
Brownian particle $A$, since the growth $L^*\propto t^{1/3}$ now
suppresses such finite size contributions for any given density.

A number of questions and opportunities for further work arise from
our analysis. In our analysis (as in that of Bray and
Blythe~\cite{BrayBlythePRL,BlytheBrayPRE}) we have simply applied the
Pascal principle that the best survival strategy of a particle in a sea
of moving traps is to remain stationary.  While it seems intuitively
obvious that this principle would hold whether the motions are diffusive
or subdiffusive, the proof has only been presented for the diffusive
case~\cite{BrayMajumBlythePRE,MoreauEtAlCondMat}. As an interesting
aside, we note that while the proof of the Pascal principle has assumed
an equal concentration of traps on either side of the particle, we
conjecture that this is not a necessary condition, and that the
Pascal principle also holds with an asymmetric distribution and even if the
traps are all located on one side of the particle.  This, too, remains
to be proved.

At this point we inject a digression that is relevant not
only to our analysis but also to the original work of Bray and
Blythe~\cite{BrayBlythePRL,BlytheBrayPRE}.
They assumed that the particle $A$ is initially surrounded by a random
(Poisson) distribution of mobile traps, an assumption also made in our
analysis, cf. Eq.~\eqref{notraps}.  On the other hand, if at the start of
the observations ($t=0$) the process has already been taking place for
some time $-\tau$ (i.e., if the process started at some time $\tau$ in
the past), then it is known that the distribution around the
surviving particles at time $t=0$ is not of Poisson form.  Those
particles that initially had close-by traps are more likely to have
been trapped already than those that did not, so that those particles
that have survived are surrounded by a region of
fewer than average traps (sometimes referred to as a ``gap").
Bramson and Lebowitz arrive at the conclusion that the configuration of
$B$ particles is nevertheless dominated by a Poisson random
measure~\cite{BramsonLebowitz2}.  In Appendix~\ref{sec:initial} we confirm
that for any finite $\tau$ the gap does not affect the asymptotic survival
probability results of Bray and Blythe.  The detailed nature of the
gap is different in the diffusive and subdiffusive cases, and unknown in
the latter. However, we
conjecture that it is no more pronouced in the subdiffusive than in the
diffusive system, and that it does not affect our results either.

Our own results of course leave a number of questions unanswered.  One
obvious question concerns the marginal role of the trap
exponent $\gamma=2/3$ when
the particle is diffusive.  Why is this a marginal exponent?  A
connection between this critical value and the fact that for a Brownian
particle the length that maximizes the lower bound of the survival
probability grows as $L^*\sim t^{1/3}$ seems plausible, but the
conceptual basis for such a relation is not clear.

The most pressing and intriguing puzzle to resolve is that of calculating
the survival probability when the particle $A$ is diffusive
($\gamma'=1$) and the traps are extremely subdiffusive ($0<\gamma<2/3$).
Because the upper and lower bounds in this case do not have the same
asymptotic time dependence, we are not able to say anything about this
case on the basis of the procedures used in this paper.

\appendix
\section{MODIFIED INITIAL CONDITION}
\label{sec:initial}
Suppose that the process $A+B \to B$ with a diffusive particle and
diffusive traps began at time $-\tau$, but our observation of the system
starts at time $t=0$.  Even if at time $-\tau$ the distribution of
$B$'s around $A$ was random (Poisson), it will not be so at time $t=0$.
Let $p(r)$ be the probability density of finding an empty region of
length $r$ to the right of $A$ at time $t=0$, so that the first trap
is found to be located between $r$ and $r+dr$.  This function is unknown
for our system, but one can conjecture a behavior on the basis of known
(analytic and numerical) approximate results for the bimolecular reactions
$A+A\to \emptyset$ and $A+B\to
\emptyset$~\cite{ArgyrakisKopelmanI,ArgyrakisKopelmanII},
\begin{subequations}
\label{prt}
\begin{align}
\label{prtsmall}
p(r)&\sim r/\langle r \rangle \qquad \qquad \text{for} \quad  r/\langle
r \rangle \ll 1 \\
p(r)&\sim \exp(-r/\langle r \rangle)\quad \text{for} \quad r/\langle r
\rangle \gg 1
\label{prtlarge}
\end{align}
\end{subequations}
where $\langle r\rangle$ is the mean size of the gap next to particle
$A$ at the initial time of observation.
Next we need to define the function $q(r)$, the probability density that
a particle $B$ is located at a distance $r$ from $A$ at $t=0$.
This function can be related to $p(r)$, most easily by considering a
discrete lattice and then going to the continuum limit. Let
$p_n$ be the probability of a gap of size $n$ and a particle $B$ at
position $n+1$ to the right of $A$.  Thus
$p_n=\prod_i^n (1-q_i) q_i$, where $q_i$ is the probability that there
is a trap $B$ at distance $i$ from $A$. It then follows that
$\ln p_n=\ln(q_i)+ \sum_i^n \ln(1-q_i)\approx \ln(q_i)-\sum_i^n q_i$
and thus in the continuum limit
\begin{equation}\label{}
p(r)=q(r)\exp\left[-\int_0^r q(x) dx \right].
\end{equation}
To find $q(r)$ one thus needs to solve a nonlinear integral equation, a
difficult task not worth the attempt given our imperfect
knowledge of $p(r)$.  It is
nevertheless straightforward to conclude that the solution
$q(r)=p(r)$ for $r\ll \langle r\rangle$ is compatible
with \eqref{prtsmall}, and that $q(r)=$const. for $r\gg \langle
r\rangle$ is compatible with \eqref{prtlarge}.  For our purposes it is
thus sufficient to approximate $q(r)$ by
\begin{align}\label{qrL}
q(r)=\begin{cases}
\dfrac{1}{R} \dfrac{r}{\langle r\rangle}, \quad 0\leq r\le \langle
r\rangle,\\       \dfrac{1}{R}, \quad\quad\quad r\ge \langle r\rangle,
   \end{cases}
\quad \text{for}\quad {R\toi}.
\end{align}
The coefficient $1/R$ has been set by requiring that the integral of
$q(r)$ from zero to $R$ be normalized to unity.
Thus, the presence of $A$ affects the distribution of $B$'s within a
distance $\langle r \rangle$ from $A$ but not beyond, where the
distribution remains random.

How does this deviation from a random distribution affect the survival
probability? For this calculation we also introduce
$q(r;R)=q(r)/\int_0^R q(r)dr$, the probability density for a
$B$ particle to be found at a distance $r$ from $A$ given that there is
a $B$ particle somewhere in an interval of length $R$ that begins at $A$.
An upper bound is obtained, as before, by invoking the Pascal principle.
Generalizing the derivation for a Poisson distribution one
straightforwardly arrives at
the expression
\begin{align}
Q_\infty(t)
&=\lim_{R\to\infty} \left[\int_0^R q(r;R) Q_1(t|r) dr\right]^{\rho
R}\nonumber\\
&=\lim_{R\to\infty} \left[\int_0^R q(r;R) -\int_0^R q(r;R) \widehat
Q_1(t|r) dr\right]^{\rho R},
\end{align}
where $Q_1$ is given in Eq.~\eqref{Brownian} and $\widehat Q_1 \equiv 1-Q_1$.
Since $\lim_{R\to\infty} \int_0^R q(r;R)dr =1$, it follows that
\begin{align}
Q_\infty(t)&=\lim_{R\to\infty} \exp\left[ -\rho R \int_0^R q(r;R)
\widehat Q_1(t|r) dr\right]
=\lim_{R\to\infty} \exp\left[ -\frac{\rho R}{\int_0^R q(r)dr} \int_0^R
q(r)\widehat Q_1(t|r) dr\right]\nonumber\\
&=\lim_{R\to\infty} \exp\left[ -\rho R\int_0^R q(r)\widehat Q_1(t|r)
dr\right]\nonumber\\
&=\exp\left[ -\rho \left( \int_0^{\langle r\rangle}
\frac{r}{\langle r\rangle}\widehat Q_1(t|r) dr +\int_{\langle
r\rangle}^\infty \widehat Q_1(t|r) dr \right)\right],
\label{Qinfty}
\end{align}
where we have introduced the explicit form \eqref{qrL} in the last line.
The integrals can be carried out in the long time limit, to yield
\begin{equation}\label{}
Q_\infty(t)=\exp\left[ -\rho\left(\frac{\langle
r\rangle}{2}+\sqrt{\frac{4Dt}{\pi}}\right) \right]  \quad
\text{for}\quad t\toi.
\end{equation}
Since $\langle r \rangle$ is a constant (no matter how large),
eventually the second term in the exponent dominates, and the upper
bound on the survival probability in the diffusive case is exactly the
same as obtained for an initially Poisson distribution of $B$'s around $A$.

Consider now the lower bound on the survival probability. The
probabilities $Q_2$ and $Q_3$ are not changed by a modification of the
initial distribution of traps around the particle.  The probability
$Q_1$ that at $t=0$ a box of size $L$ around $A$ contains no traps is
changed.  It is again easiest to start with a discrete lattice
and define $\bar
\omega_n$ as the probability that there is a gap of $n$ sites to the right of
$A$ given that there is a single trap to the right of $A$ as determined
by the distribution $q_i$.  Thus
\begin{equation}
\bar \omega_n=\prod_{i=0}^n (1-q_i)\Rightarrow \ln \bar\omega_n
=\sum_{i=0}^n \ln(1-q_i)\approx -\sum_{i=0}^n q_i.
\end{equation}
The continuum limit gives
\begin{equation}\label{}
\bar \omega(x)=\exp\left( -\int_0^x q(r) dr \right).
\end{equation}
If there are $N=\rho R$ to the right of $A$, with
$R\toi$, the probability of a gap of size $x$ to the right of $A$ is
$\omega(x)=\bar\omega^{N}(x)=\bar\omega^{\rho R}(x)$.  Inserting the
expression for $q(r)$ and carrying out the integrals for $R\toi$ gives
\begin{equation}\label{}
\omega(x)=\begin{cases}
\displaystyle \exp\left[-\rho x^2/(2 \langle r\rangle) \right], &  x\le
\langle r\rangle\\[0.3cm]
\displaystyle \exp\left[-\rho (x-\langle r\rangle/2)\right],  &  x\ge
\langle r\rangle
\end{cases}
\end{equation}
The same result is valid for a gap to the left of $A$, so that
the probability $Q_1$ for there to be a gap of length
$L$ centered at $A$ then is
\begin{align}\label{}
Q_1=&\displaystyle \exp\left[-\rho L^2/(4\langle
r\rangle)\right],  \qquad \text{for}\quad  L/2\le \langle r\rangle
\label{P1bad}\\[0.3cm]
Q_1=&\displaystyle \exp\left[-\rho (L-\langle
r\rangle)\right],  \qquad  \text{for}\quad   L/2\ge \langle r\rangle
\label{P1good}
\end{align}
The lower bound on the survival probability then is
\begin{equation}
P_L(t)=Q_1\frac{4}{\pi} \exp\left(-\frac{\pi^2 D' t}{l^2}-2
\rho\sqrt{\frac{4Dt}{\pi}}\right),
\end{equation}
where we have inserted the explicit expressions for $Q_2$ and $Q_3$
found earlier.
As before, one can choose the box size $L$ to maximize this lower bound.
There are now two possibilities.  One is to assume that $L\leq 2\langle
r\rangle$, whence
\begin{align}\label{ptc1}
P_L(t)=\frac{4}{\pi} \exp\left(-\frac{\pi^2 D' t}{L^2}-\rho
\frac{4L^2}{\langle r\rangle} -2 \rho\sqrt{\frac{4Dt}{\pi}}\right).
\end{align}
Maximizing with respect to $L$ gives
$L^*=\left(4 \pi^2 D' \langle r\rangle t/\rho\right)^{1/4}$, which grows
with time and is therefore in contradiction with the assumption $L\leq
2\langle r\rangle$.  Thus we have to take the other possibility,
$L\ge 2\langle r\rangle$, whence Eq.~\eqref{P1good} leads to
\begin{align}\label{ptc2}
P_L(t)=\frac{4}{\pi} \exp\left(-\frac{\pi^2 D' t}{L^2}-\rho L
+\rho \langle r\rangle-2 \rho\sqrt{\frac{4Dt}{\pi}}\right),
\end{align}
and maximizing with respect to $L$ gives
$L^*=\left(2\pi^2 D' t/\rho\right)^{1/3}$.
For $t \toi$, the new term $\rho \langle r\rangle$ in the exponent is
negligible compared to the terms that grow with time.  The asymptotic
lower bound is therefore identical to that obtained by Bray and Blythe
for the Poisson distribution of traps around $A$.

Although the distribution $q(r)$ is not known in the subdiffusive case,
it will surely be the case that at large distances from $A$ it is
independent of $r$ even if the trapping process has been going on for
a long time $\tau$ before the start of the observations at $t=0$.
Therefore in this case as
well the asymptotic survival probability results obtained in this paper
are applicable.

\acknowledgments     %>>>> equivalent to \section*{ACKNOWLEDGMENTS}
This work was partially supported by the Ministerio de Ciencia y
Tecnolog\'{\i}a (Spain) through Grant No. FIS2004-01399, and by the
National Science Foundation under grant No. PHY-0354937.

%\bibliography{report}   %>>>> bibliography data in report.bib
%\bibliographystyle{spiebib}   %>>>> makes bibtex use spiebib.bst

\end{document}